\documentclass[10pt]{iopart}
\usepackage{epsf}
\usepackage{times}
\usepackage{harvard}

\newcommand{\notext}[1]{}

\newcommand{\half}{{\textstyle\frac12}}
\newcommand{\bg}[1]{\mbox{\mathversion{bold}$#1$\mathversion{normal}}}

\newcommand{\be}{\begin{equation}}
\newcommand{\ee}[1]{\label{#1}\end{equation}}
\newcommand{\bes}{\numparts\begin{eqnarray}}
\newcommand{\ees}[1]{\label{#1}\end{eqnarray}\endnumparts}

\newcommand{\bss}[1]{{\mbox{\sffamily\bfseries #1}}}

\begin{document}
\ifx\pdfoutput\undefined \newcount\pdfoutput \fi
\pdfoutput=0
\jl{3}
\title[The metal--insulator transition]{The metal--insulator transition in disordered systems:\\
a new approach to the critical behaviour}
\author{Alex P Taylor\footnote{Present Address: The Judge Institute of Management,
University of Cambridge, Trumpington Street, Cambridge CB2~1AG, U.K.} 
and Angus MacKinnon}
\address{Blackett Laboratory, Imperial College, London SW7~2BW, U.K.}
\bibliographystyle{JPhysics}
\begin{abstract}
In the most popular approach to the numerical study of the Anderson
metal--insulator transition the transfer matrix method is combined 
with finite--size scaling ideas.  This approach requires large computer 
resources to overcome the statistical fluctuations and to accumulate data
for a sufficient range of different values of disorder or energy.  
In this paper we present an alternative approach in which the basic 
transfer matrix is extended to calculate the derivative with respect to disorder.
By so doing we are able to concentrate on a single value of energy or disorder 
and, potentially, to calculate the critical behaviour much more efficiently 
and independently of the assumed range of the critical regime.  
We present some initial results which illustrate both the advantages 
and the drawbacks of the method.
\end{abstract}
\submitted

\section{Introduction\label{chaptmsnow}} 
The \citeasnoun{And58} transition in three--dimensional
electronic systems has been extensively studied over recent years \cite{KmK93}. 
It is
well known that increasing the fluctuations of the random potential
in three--dimensional systems (3D) causes a transition from metal to
insulator. This transition occurs due to a change in the nature of the
electron wavefunctions under the influence of the disorder. In the
insulating phase the magnitude of the potential fluctuations is
sufficiently large to localise the wavefunctions to a region of
space. As the disorder is reduced below a critical value the electron
states become delocalised allowing conduction through the system. In
models of disordered systems such as the Anderson Hamiltonian this
change can be measured numerically by observing the change in the decay
length of the transmission probability (the correlation length $\xi$)
using the transfer matrix method \cite{mKK81,macK94}. Since the
development \cite{AALR79} of a scaling theory for the zero--temperature
dc conductance of disordered electronic systems much numerical work has
focused on calculating the critical exponent for the correlation length
\cite{mKK81,macK94,SO97,SO99}. 

The assumption of one--parameter scaling
implies that the renormalised length $\Lambda_1 =\xi_1/M$ should follow the 
scaling equation:
\be
 {\rmd\ln  \Lambda_1 \over \rmd\ln M}=\chi_1 (\ln \Lambda_1 )
\ee{equ:standards}
with solutions of the form $\Lambda_1 = f _1(M/\xi_{\infty} )$.
When plotted against disorder, $W$, for different sizes, $M$, the
curves, $f_1$, all cross at a fixed point $\chi_1 = 0$ corresponding to
the metal--insulator transition. $\xi_{\infty}$ is the correlation
length  for the infinite system which diverges close to the critical point,
$W_{\rm c}$, as $\xi_{\infty} \sim \mid W-W_{\rm c} \mid ^{-\nu}$.
Near the critical point the scaling equation can be linearised:
\be
\Lambda_1=\Lambda_{1,{\rm c}}+a(W-W_{\rm c})M^{1\over \nu} . 
\ee{equ:critlam}
The standard method of calculating the critical exponent $\nu$ is
to analyse the finite size scaling in terms of this equation.
However, taking the logarithm of the derivative with respect to disorder
gives us:
\be
\ln\left(\left.{\rmd\Lambda_1 \over \rmd W}\right |_{W_{\rm c}}\right)
=\ln a + {1\over \nu} \ln M . 
\ee{equ:fit}
This expression provides a way of calculating the critical
exponent from the derivative of the correlation length of the
finite system $\Lambda_1$. In previous work
\cite{macK90a} it has been noted that the
calculation of the derivative ${\rmd\Lambda_1 / \rmd W}$ would
provide useful complimentary information about the scaling and a
more direct approach to calculating the critical exponent. However
direct estimation of the derivative has up to now been plagued by
large numerical errors associated with numerical division 
when using a finite difference approximation $[\Lambda_1(W+\Delta
W)-\Lambda_1(W)]/\Delta W$. 

Our new method is an adaptation of the
standard transfer matrix approach to calculate the derivative, $\bss{\.T}$, of
the transfer matrix, $\bss{T}$, from which it is possible to
calculate the expectation value of $\rmd\Lambda_i /\rmd W$, and hence
find $\nu$. This approach is applicable for 3D cubic systems
(M=L=6,8,10,12,14) rather than for quasi--1D systems. In contrast
to quasi--1D systems, where the conductance is described by only one
correlation length, the conductance of cubic systems is given by a
sum over contribution from a series of correlation lengths. Here we
present preliminary results for the scaling of each correlation
length individually as well as for the conductance as a whole.

\section{Computational  Method\label{meths}}
\subsection{The traditional transfer matrix\label{sect:tranmat}}
The \citeasnoun{And58} model for non--interacting electrons on a simple cubic
lattice in zero magnetic field is described by the Hamiltonian:
\be
 H = \sum_i \epsilon_i |i \rangle \langle i|
+ \sum_{i \neq j} V |i \rangle \langle j|, 
\ee{anderson}
where $|i \rangle$ is the atomic orbital at site $i$, and $V=1$ for
nearest neighbours and zero otherwise. The disorder is introduced
by allowing the on--site energies $\epsilon_i$ to behave stochastically.
The site energies $\epsilon_i$ are chosen according to the probability
distribution
\be
 p(\epsilon) = W^{-1}\theta (\half W - |\epsilon|),
\ee{box}
where $\theta$ is the step function so that $p(\epsilon)$ is a box
distribution. The Schr\"odinger equation $H \phi = E \phi$ for a
system with cross section $M\times M$ and length $L$ can be rewritten in
recursive form:
\be
   \left(\begin{array}{l}\bg{\psi}_{i+1} \\ \bg{\psi}_{i}\end{array}\right) 
= \left(\begin{array}{cc} E-\bss{H}_i & -\bss{1} \\ \bss{1} & \bss{0}\end{array}\right)  
\left(\begin{array}{l}\bg{\psi}_{i} \\ \bg{\psi}_{i-1} \end{array}\right)
= \bss{T}_i \left(\begin{array}{l}\bg{\psi}_{i} \\ \bg{\psi}_{i-1}\end{array}\right),
\ee{equ:stak}
where the $2M^2\times2M^2$ matrix, $\bss{T}$, is the so-called the transfer
matrix and the $M^2\times M^2$ matrix, $\bss{H}_i$ is the Hamiltonian 
of a single slice disconnected from the rest of the lattice. 
The application of $\bss{T}_i$ to a slice gives the solution of the Schr{\"o}dinger
equation for the next slice. In this way the complete wavefunction
for a disordered sample of length $L$ can be generated by repeated
application of the transfer matrix:
\be
   \left(\begin{array}{l}\bg{\psi}_{L+1} \\ \bg{\psi}_{L}\end{array}\right)  
= \prod_{i=1}^L \bss{T}_i \left(\begin{array}{l}\bg{\psi}_{1} \\ \bg{\psi}_{0}\end{array}\right) .
\ee{equ:TM}
The standard transfer matrix method then consists of calculating
the asymptotic (\mbox{i.e.} $L\rightarrow\infty$) behaviour of the eigenvalues of the Hermitian matrix $\bss{Q}$:
\be
\bss{Q}=\bss{T}^{\dagger}\bss{T}
\ee{equ:Qdef}
by repeated orthogonalisation of the columns of $\bss{T}$  \cite{mKK83a}.  Alternatively, for a finite $L$, 
a system in  \citeasnoun{Lan70} geometry, \mbox{i.e.} embedded between perfectly ordered  leads, 
may be represented in the form
\be
\bss{Q}= \bss{U}^{\dagger}\left[\prod_{i=1}^L\bss{T}_i\right]^{\dagger}
\bg{\Lambda}\bss{U}^*\bss{U}^{\mathrm T}\bg{\Lambda}
\left[\prod_{i=1}^L\bss{T}_i\right]\bss{U}
\ee{equ:Qdef2}
where the columns (rows) of $\bss{U}$ ($\bss{U}^T\bg{\Lambda}$) are the right (left) eigenvectors which
diagonalise $\bss{T}$ in the absence of disorder and 
\be
\bg{\Lambda}=\left(\begin{array}{cc}\bss{1}&\bss{0}\\ \bss{0}&-\bss{1}\end{array}\right) .
\ee{equ:defLambda} 
The correlation lengths $\xi_i$, from
which the scaling properties are calculated, are related to the 
eigenvectors $|z_i>$ of $\bss{Q}$  by
\be
\bss{Q}|z_i>=\exp(z_i) |z_i>\qquad\mbox{where}\quad z_i = 2\alpha_i L = 2L/\xi_i
\ee{equ:lyan}
for $i=1,2,...2M^2$. 
In the quasi 1D limit the $\alpha_i$ are called Lyapunov
exponents. 

The normalised eigenvectors for the system with zero disorder can
be grouped together to form a $2M^2 \times 2M^2$ matrix, $\bss{U}$, 
in which first $M^2$
columns are made from the $2M^2\times M^2$ sub--matrix ($\bss{U}_{+}$) of
eigenvectors representing waves travelling in the positive
z--direction, and the last $M^2$ columns from those travelling in the opposite
direction, ($\bss{U}_{-}$). The set of left--handed eigenvectors is given by
$\bss{U}^{\mathrm T} \bg{\Lambda}$. The eigenvectors can be used to transform the real--space
transfer matrix \eref{equ:stak} into the wave representation:
\be
   \bss{T}_{\mathrm w}=\bss{U}^{\mathrm T} \bg{\Lambda} \bss{T}_L \bss{U},
\ee{equ:transT}
where $\bss{T}_L$ is the real--space transfer matrix for a system of
length $L$. The transfer matrix in the wave representation, $\bss{T}_{\mathrm w}$,
can be expressed in terms of the transmission and reflection
matrices:
\be
   \bss{T}_{\mathrm w}= \left(\begin{array}{cc}
                \bss{t}-\bss{r}'(\bss{t}')^{-1}\bss{r} & \bss{r}'(\bss{t}')^{-1}\\ 
                -(\bss{t}')^{-1}\bss{r} & (\bss{t}')^{-1} 
                \end{array}\right). 
\ee{tranmat}
This suggests that $\bss{r}'$ and $\bss{t}'$ may be calculated from the matrix
products:
\bes
   \bss{t}'^{-1} &=& \left(\begin{array}{cc} \bss{0} & \bss{1}
                        \end{array}\right) \bss{T}_{w} 
                        \left(\begin{array}{c} \bss{0} \\ \bss{1} 
                        \end{array}\right)
 =\bss{U}_{-}^T  \bg{\Lambda} \bss{T}_L \bss{U}_{-} =\bss{U}_{-}^T  \bg{\Lambda} 
\prod_{n=1}^L \bss{T}_n \bss{U}_{-}
\label{wep}\\
   \bss{r}'\bss{t}'^{-1} &=&  \left(\begin{array}{cc} \bss{1} & \bss{0}
                                \end{array}\right)\bss{T}_{w} 
                              \left(\begin{array}{c}\bss{0} \\ \bss{1}
                                \end{array}\right)      
 = \bss{U}_{+}^T \bg{\Lambda} \bss{T}_L \bss{U}_{-} 
 = \bss{U}_{+}^T \bg{\Lambda} \prod_{n=1}^L \bss{T}_n \bss{U}_{-} .
\ees{rep}

\subsection{Stabilisation of the Iteration\label{sect:stable}} 
In practice the repeated multiplications of the matrix $\bss{U}_{-}$ by
successive transfer matrices gives rise to numerical instabilities
that require attention. The matrix gradually becomes
dominated by the largest eigenvalues and the smaller ones are lost
owing to the finite accuracy of the numerical process. It is these
eigenvalues that are needed when the inverse is taken to calculate
$\bss{t}'$. To avoid losing them the algorithm must be modified. After
about ten iterations, and well before the small
eigenvalues are lost, the relevant information is `stored' by
multiplying from the right by a stabilising matrix, $\bss{Y}$:
\be
   \prod_{n=1}^{10} \bss{T}_n \bss{U}_{-} \bss{Y} = \bss{Z}\bss{Y}
\ee{equ:stable}
The matrix $\bss{Y}$ may be chosen to be the inverse of the top half 
of the $\bss{Z}$ (or the upper triangular matrix which orthonormalises the columns of $\bss{Z}$):
\be
   \bss{Z}\bss{Y} = \left(\begin{array}{c} \bss{Y}^{-1} \\ \bss{G}
                            \end{array}\right)
        \bss{Y} = \left(\begin{array}{c} \bss{1} \\ \bss{G}\bss{Y}
                            \end{array}\right) .
\ee{equ:maket}
This process of multiplying by the inverse matrix is repeated
approximately every ten iterations. The product of the $\bss{Y}$ is
stored separately and is used at the end to solve for the
reflection/transmission coefficients. Applying this method to
equations \eref{wep} and \eref{rep} results in the following
relations:
\bes
   \bss{t}'^{-1}\bss{Y}=\bss{A},
\label{A}\\
   \bss{r}'\bss{t}'^{-1}\bss{Y}=\bss{B},
\ees{B}
where the matrices $\bss{A}$, $\bss{B}$ and $\bss{Y}$ are known.
The numerically stable solutions are:
\bes
   \bss{t}'=\bss{Y}\bss{A}^{-1},
\label{At}\\
   \bss{r}'=\bss{B}\bss{A}^{-1},
\ees{Bt}

\subsection{Calculating the derivatives\label{meths:deriv}} 
Our new approach extends the transfer matrix method to
calculate $\bss{\.T}=\rmd\bss{T}/\rmd W$ from which the ${\rmd \xi_i / \rmd W}$ can
then be found. To preserve the multiplicative behaviour a larger
matrix $\bss{K}$ is used:
\be
\bss{K}_L=\left(\begin{array}{cc}\bss{T}_L & \bss{0} \\\bss{\.T}_L & \bss{T}_L\end{array}\right)
=\prod_{i=1}^L\left(\begin{array}{cc}\bss{T}_i & \bss{0} \\\bss{\.T}_i & \bss{T}_i\end{array}\right) 
\ee{equ:super}
where $\bss{\.T}_i$ has the simple form:
\be
\bss{\.T}_i=\left(\begin{array}{cc}\bss{\.H}_i & \bss{0} \\ \bss{0} & \bss{0} \end{array}\right) 
\ee{equ:b}
where $\bss{\.H}_i$ is the derivative of the Hamiltonian of slice $i$:
\be
\bss{\.H}_i ={\rmd\bss{H} \over \rmd W}= \left(\begin{array}{ccccc}     
                \epsilon'_1     & 0             & \ldots        & 0                     & 0     \\
                0               & \epsilon'_2   & \ldots        & 0                     & 0     \\
                \vdots          & \vdots        & \ldots                & \vdots                & \vdots\\
                0               & 0             & \ldots        & \epsilon'_{M^2-1}     & 0     \\
                0               & 0             & \ldots        & 0                     & \epsilon'_{M^2}
        \end{array}\right)
\ee{equ:c}
where $\epsilon_i'$ is a random number between $-0.5$ and $0.5$ from
\eref{box}. The larger size of the matrix increases the
computational time required but because of the additional
information contained in the derivative only one calculation per
system size is needed to calculate $\nu$, rather than many
calculations over a wide range of disorders as in the standard
method. The computational details of how to calculate the
derivatives of $g$ and $\xi_i$ from $\bss{\.T}_i$ are described in \sref{meths}. 
In \sref{sectg} preliminary results of the scaling of
these quantities is presented. 

This method cannot be applied to long systems (\mbox{i.e.} quasi--1D)
because, from Osledec's theorem, we would obtain a limiting $\bss{K}_L$
matrix of eigenvalues which cannot be related back to the
eigenvalues of $\bss{T}$. A possible solution would be to
stabilise  $\bss{T}$ and $\bss{\.T}$ simultaneously so that they would both
lead to limiting matrices of eigenvalues. However, two different
sets of vectors are required to orthogonalise $\bss{T}$ and $\bss{\.T}$ and
this creates problems at the end of the calculation when it
becomes necessary to remove the stored stabilisation vectors.  The
only way to overcome these difficulties is to consider small cubic
systems which do not need stabilising. In this work cubic systems
of size $M = 6,8,10,12$ were used. 

In the polar decomposition the transfer matrix can be parameterised 
as \cite{MPK88}:
\begin{eqnarray}
    \bss{T} &=& \bss{U}\bg{\Gamma}\bss{V} ,\\
        &=&\left(\begin{array}{cc}\bss{u}_4 & \bss{0}\\ \bss{0} & \bss{u}_2^{\dagger} \end{array}\right)
          \left(\begin{array}{cc}(\bg1+\bg{\lambda})^{1/2}&\bg{\lambda}^{1/2}\\ \bg{\lambda}^{1/2}&(\bg1+\bg{\lambda})^{1/2}\end{array}\right)
      \left(\begin{array}{cc}\bss{u}_1 & \bss{0}\\ \bss{0} & \bss{u}_3^{\dagger} \end{array}\right)
\label{polar}
\end{eqnarray}  
where $\bss{u}_i$ are arbitrary $M \times M$ unitary matrices and $\bg{\lambda}$ 
is a real diagonal matrix with positive elements $\{\bg{\lambda}_i \} $. 
If time-reversal symmetry holds then 
$\bss{u}_3=\bss{u}_1^{T}$ and $\bss{u}_4=\bss{u}_2^{T}$. In this parameterisation the transmission 
matrix $\bss{t}'$ can be written as:
\begin{equation}
\bss{t}'=\bss{u}_3(\bg{1}+\bg{\lambda})^{-1/2}\bss{u}_2
   \label{ rt}
\end{equation} 
and the conductance as:
\begin{eqnarray}
g&=&\Tr(\bss{t}'\bss{t}'^{\dagger})=\Tr(\bss{u}_3(\bg1+\bg\lambda)^{-1}\bss{u}_3^{\dagger}),\\
 &=&\sum_{i=1}^M {1\over 1+\lambda_i}
   \label{conductance}
\end{eqnarray}
The Lyapunov exponents for the ensemble are derived by substitution of the polar decomposition (\ref{polar}) 
in the matrix product $\bss{Q}$ (\ref{equ:Qdef},\ref{equ:Qdef2})
\begin{equation}
Q = \pmatrix{\bss{u}_1(\bg{1}+2\bg{\lambda})\bss{u}_1^{\dagger}&
2\bss{u}_1 \sqrt{\bg{\lambda}(\bg{1}+\bg{\lambda})}\bss{u}_3\cr
2\bss{u}_3^{\dagger} \sqrt{\bg\lambda(\bg1+\bg\lambda)} \bss{u}_1^{\dagger}&
\bss{u}_3^{\dagger}(\bg1+2\bg\lambda)\bss{u}_3 \cr}
\label{equ:Qpolar}
\end{equation} 
$\bss{Q}$ is Hermitian positive so the eigenvalues are positive real numbers and 
the flux conservation constraint, $\bss{Q}^{-1}=\bg{\Lambda}\bss{Q}\bg{\Lambda}$  
requires that the eigenvalues come in inverse pairs.
Strictly speaking the Lyapunov exponents $\alpha_i$ (\ref{equ:lyan}) are defined only in the 
quasi-1D limit, but in this work we will also use the term for arbitrary $L$.
Applying Osledec's theorem of random matrix products  \cite{PS81a,PS81b}, the Lyapunov
exponents are self--averaging for large $L$.  Hence, taking the limit of large $L$ is 
statistically equivalent to an ensemble average over many different 
realisations of the disorder. 
Computationally, for samples of width $M$ close to the transition, convergence of the first 
Lyapunov exponent  to $1\%$ accuracy is achieved for a length $L \sim 10^4 \times M$.
The conductance can be expressed in terms of
the Lyapunov exponents \cite{Pic84}:
\begin{equation}
 g=2\sum_{i=1}^N {1\over 1+\cosh(2L\alpha_i)} 
\label{equ:gexp}
\end{equation} 
In the quasi-1D limit ($L \gg 1$) the conductance can be written as:
\begin{equation}
 g\sim \sum_{i=1}^N e^{-2\alpha_i L}= e^{-2\alpha_1L} \left(1 +\sum_{i=2}^N e^{-2( 
\alpha_i -\alpha_1)}\right)
\label{equ:limit}
\end{equation} 
where the $N\sim M^{d-1}$ Lyapunov exponents are ordered and $\alpha _1$
is the smallest. As the length of the system is increased the fractional 
contribution to the current from the higher Lyapunov exponents rapidly decreases.
The conductance is dominated by the first Lyapunov exponent and the system 
is similar to the 1D case where there is only one Lyapunov exponent (one conductance 
channel) and all the states are localised.   
Since there are no other relevant length scales present the localisation
length is equivalent to the correlation length of the finite system, $\xi_1$, 
and is defined as:
\begin{equation}
{1\over \xi_1}=-\lim_{L\rightarrow \infty}
{\ln g(L)\over 2L}=\lim_{L\rightarrow \infty} \alpha_1
\label{equ:}
\end{equation}

From the Hellman--Feynman theorem the expectation value of the derivative of the
eigenvalues, $\tau_i$,  of $\bss{t}'^\dagger\bss{t}'$ can be calculated:
\bes
{\rmd\tau_i \over \rmd W}  &=& \bi{u}_i^\dagger {\rmd\bss{t}'^{\dagger} \bss{t}'
\over \rmd W} \bi{u}_i \\
 &=& 2\Re\left(\bi{u}_i^\dagger \bss{t}'^{\dagger}{\rmd\bss{t}' \over\rmd W} \bi{u}_i 
\right) ,
\ees{equ:start}%
where $\bi{u}_i$ is the eigenvector corresponding to $\tau_i$.
To find the $\bi{u}_i$  and $\tau_i$ it
is necessary to perform an initial calculation of the transfer
matrix as outlined in \sref{sect:tranmat}.  Then $\bi{u}_i$ is used in
a calculation of the derivative from equation \eref{equ:start}. The
expectation value of the derivative is only calculated for the
smallest ten Lyapunov exponents to avoid the necessity of stabilising
the matrices. If the expectation value of the whole matrix was to
be calculated, the smallest expectation values would rapidly
become insignificant compared to the largest, and would be lost in
the rounding errors.  In general the expectation value for a quasi
1D system cannot be calculated because as $L$ increases the result
rapidly becomes independent of initial conditions, i.e.
independent of the initial vector $\phi_i$.

Starting from equation \eref{equ:start} we derive the form used in
the computational calculations. From \eref{wep} we have:
\be
\bss{t}'^{-1}=\bss{U}_{-}^{\mathrm T}\bg{\Lambda}\bss{T}_L\bss{U}_{-} 
\ee{equ:f}
where $\bss{U}_{-}$ correspond to eigenvectors in the leads with negative $k_z$.
Taking derivatives and using equation \eref{equ:super} we obtain:
\be
{\rmd\bss{t}'^{-1}\over\rmd W}
\ee{equ:g}
from the identity $\bss{t}'^{-1}\bss{t}'=1$ we can write:
\be
{\rmd\bss{t}'\over\rmd W}=-\bss{t}'{\rmd\bss{t}'^{-1}\over\rmd W}\bss{t}' 
\ee{equ:h}
therefore we have:
\be
\bss{t}'^{\dagger}{\rmd\bss{t}' \over\rmd W}
= -\bss{t}'\bss{U}_{-}^{\mathrm T}\bg{\Lambda}{\bss{\.T}_L}\bss{U}_{-}\bss{t}' 
\ee{equ:i}
where the transmission matrices come from the initial calculation
and still contain the evanescent modes.  Only at this point can
we remove the evanescent
 modes  (indicated by application of $\bss{Q}$) and calculate the expectation value:
\be
{\rmd\over\rmd W}(1+\lambda_i)^{-1}=-\phi_i \bss{Q}^{\mathrm T} \bss{t}'\bss{U}_{-}^{\mathrm T}\bg{\Lambda}
\left(\begin{array}{c}\bss{0} \\ \bss{1} \end{array}\right)
\prod_i^L \bss{K}_i\left(\begin{array}{c}\bss{1} \\ \bss{0}
                        \end{array}\right)
        \bss{U}_{-} \bss{t}'\bss{Q} \phi_i 
\ee{equ:j} 
where $\bss{Q}$ is an $M \times M'$ matrix with all
elements zero, except for $Q_{nn} = 1,1\leq n\leq N$ which
correspond to the $N$ non--evanescent modes. The actual computation
starts with the initial vector on the right and by successive
matrix products moves through to the left. The derivative of
$\Lambda_i$ can be calculated \cite{Pic84}:
\be
{\rmd\Lambda_i \over\rmd W}=-4{\rmd\lambda_i \over\rmd W} [\cosh^{-1}(2\lambda_i
+1))^2 \sinh (\cosh^{-1}(2\lambda_i +1)) ]^{-1} 
\ee{equ:poot}
These values were found to be in agreement with the derivative as
calculated from the normal method, $\approx (\Lambda_{17.5} -
\Lambda_{16.5})$. The critical exponent is calculated by fitting a
straight line to equation \eref{equ:fit}.

The derivative of the conductance $g$ can be calculated from the
derivative of the LE's \eref{equ:poot}. The one--parameter scaling
theory is formulated in terms of the conductance and the scaling
behaviour of this quantity is considered first in the next section
before considering each LE individually.

\section{Cubic systems\label{cubics}}
\begin{figure}[htp]
        \center{\epsffile{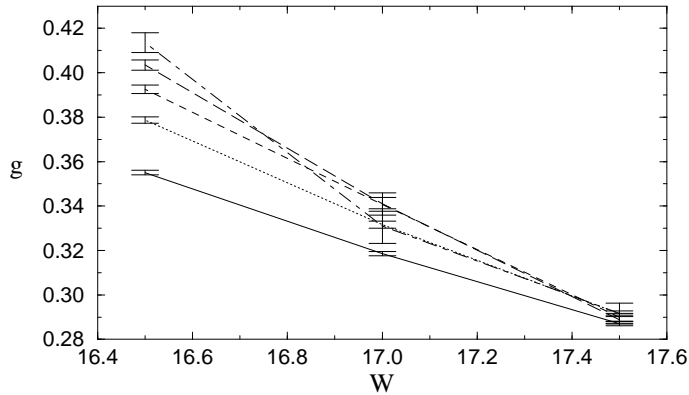}}
        \caption{ Conductance, $g$ in units of $e^2/h$, \mbox{vs.} 
disorder, $W$, for $M =  6$ (\full), $8$ (\dotted), $10$ (\broken), 
$12$ (\longbroken) \& $14$ (\chain). \label{fig:geevdis}}
\end{figure}
\subsection{Scaling of the conductance \label{sectg}}
To estimate the critical point we have plotted (\fref{fig:geevdis}) the conductance $g$
against disorder $W$ for system sizes $M=6,8,10,12,14$. The value of
critical disorder, $W_{\rm c}$,  is not expected to depend on the
quantity being considered.  Therefore, a size--independent critical
point at approximately $W=16.5$ would be anticipated to be in
agreement with the values calculated from both energy statistics
\cite{HS93,ZK94} of cubes and from the analysis of the 1st LE in
quasi--1D systems. However, it is clear that the apparent critical
disorder is above $W=16.5$ within the error bars of the
calculation. The results are also qualitatively the same for 
$\log(g)$. 
This movement of the critical point from the expected value
of $W=16.5$ is probably due to the presence of the leads connected
to the ends of the cube. This is in agreement with other results
\cite{SOK2000,KOS2000}  which find that the conductance is size dependent at
$W = 16.54$  with periodic boundary conditions. However, with hard
wall boundary conditions the conductance is found to be virtually
size--independent at  $W = 16.54$ \cite{SO97}. In energy statistics
calculations the boundary conditions affect the critical
distributions but do not appear to shift the critical point
\cite{BMP98}.

\begin{table}[htp]
\caption{At disorder $W=16.5$ the number of iterations, open modes and the 1st LE
 for cubic samples , $E=0$.\label{tab:runs}}
\begin{indented}
\item[]\begin{tabular}{@{}rrrl}
\br
M       & iterations    &modes  &\centre{1}{$\Lambda_1$}   \\
\mr 
6       &100000 &21             &$0.671 \pm  0.0015$ \\
8       &50000          &45             &$0.694 \pm  0.002$\\
10      &30000          &61             &$0.709 \pm  0.003$\\
12      &20000          &95             &$0.718 \pm 0.003$\\
14      &5000           &123            &$0.729 \pm 0.007$\\
\br
\end{tabular}
\end{indented}
\end{table}

\begin{figure}[htp]
     \center{\epsffile{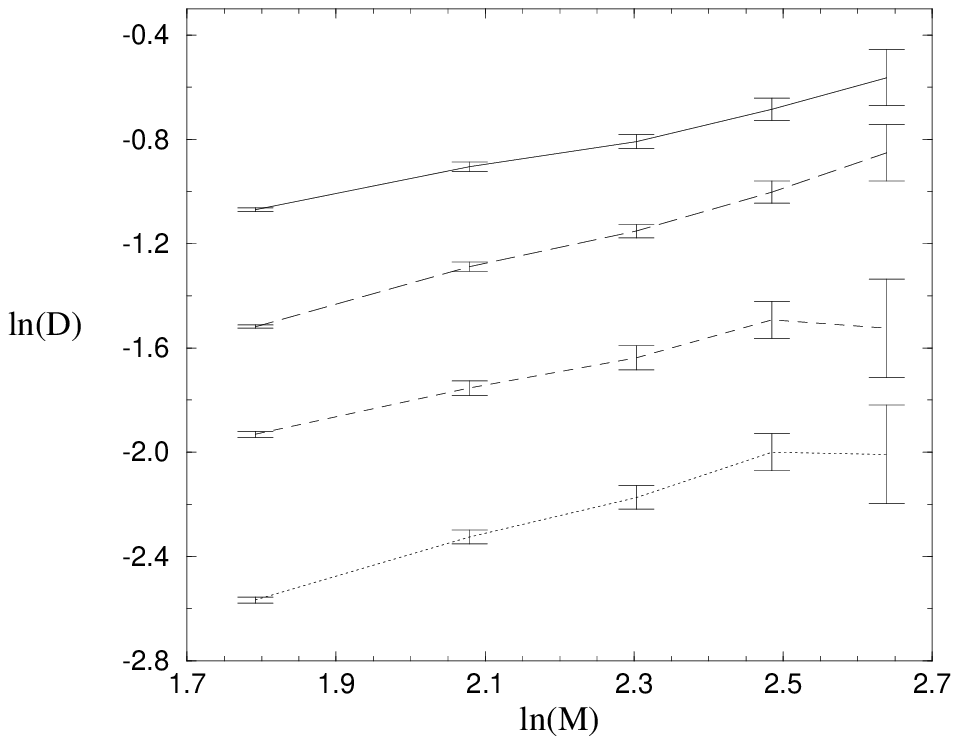}}
    \caption{$\ln D$ \mbox{vs.} $\ln M$ for $W=16.5$ and $M=6,8,10,12,14$. 
The gradient is equal to $1/\nu$. $D=\left<{\rmd\log (g)/\rmd W}\right>$ (\full),  
$\left<{\rmd\log (g)/\rmd W}\right>/\left<\log(g)\right>$ (\longbroken),  
$\left<{\rmd g/\rmd W}\right>/\left< g \right>$ (\broken), 
$\left<{\rmd g/\rmd W}\right>$ (\dotted). 
\label{fig:geescale}}
\end{figure}

\begin{table}[htp]
 \caption{Critical exponents calculated from the conductance of cubes with
sizes $M=6,8,10,12$\label{tab:geecrit}}
\begin{indented}
\item[]\begin{tabular}{@{}rclrr}
\br
W       &D                                                                      
&\centre{1}{$\nu$}      &$\chi^2$       &\centre{1}{Q}\\
\mr\medskip
16.5    &$\displaystyle{\rmd\log(g) \over\rmd W}$                               
&$1.85 \pm 0.12$        &0.7            &0.7    \\ \medskip
16.5    &$\displaystyle\left<{\rmd\log(g)\over\rmd W}\right>/\left<\log(g)\right>$            
&$1.34\pm 0.066$        &1.0            &0.6    \\ \medskip
16.5    &$\displaystyle{\rmd g\over\rmd W}$                                     
&$1.24 \pm 0.095$       &0.3            &0.9    \\ \medskip
16.5    &$\displaystyle\left<{\rmd g\over\rmd W}\right>/\left< g\right> $                     
&$1.65\pm 0.17$ &0.2            &0.9    \\ \medskip
17.5    &$\displaystyle{\rmd \log(g)\over\rmd W}$                               
&$1.34\pm 0.073$        &1.4            &0.5    \\ \medskip
17.5    &$\displaystyle\left<{\rmd\log(g)\over\rmd W}\right>/\left<\log (g)\right>$           
&$1.30\pm 0.068$        &1.2            &0.5    \\ \medskip
17.5    &$\displaystyle{\rmd g \over\rmd W}$                                    
&$1.23\pm  0.14$        &0.4            &0.8    \\ \medskip
17.5    &$\displaystyle\left<{\rmd g\over\rmd W}\right>/\left< g\right>$                      
&$1.28\pm 0.15$ &0.3            &0.9    \\
\br
\end{tabular}
 \end{indented}
 \end{table}

Applying equation \eref{equ:fit} with $g$ as the scaling parameter
gives us:
\be
\ln(D)=b+ {1\over \nu} \ln M 
\ee{equ:fit2}
with  $D= \left.\rmd g /\rmd W\right|_{W_{\rm c}}$. If $\log g$ is used
as the scaling parameter then  
$D= {1 / g}\left.\rmd g / \rmd W\right |_{W_{\rm c}}$.
Also considered were  
$D=\langle\left.\rmd g/\rmd W\right|_{W_{\rm c}}\rangle/\langle g\rangle$  
and  
$D=\langle\left.\rmd\log(g)/\rmd W\right|_{W_{\rm c}}\rangle/\langle\log(g)\rangle$. 
The plot of equation \eref{equ:fit2} is given in
\fref{fig:geescale} and the critical exponent found by a
`least--squares--fit' of the gradient is given in
\tref{tab:geecrit}. The error in the value of $g$ is considerably
smaller than the error in the gradient and can be ignored when
calculating the standard deviation of the fit.  From the error
bars on the graph we can see that $\log (g)$ produces slightly
more accurate results because it decreases the influence of the
infrequent but very large values of the derivative. 
The $M=14$ data is not used in the calculation of the critical
exponents. The results from the same calculation performed at
$W_{\rm c}=17.5$ are also given in \tref{tab:geecrit}. On average
the critical exponents appear to be slightly lower in value for
$W_{\rm c}=17.5$. Clearly the observed shift of the critical point is
completely independent of our new method to calculate the critical
exponent. However this deviation from scaling theory may well be
prejudicing the critical exponent obtained from the new method.
Nevertheless there is general agreement with previous calculations
of the critical exponent which give a value of $\nu \approx 1.5$.

There are thus two deviations from the expected results, both  of which 
are probably due to finite size effects: the shift in the apparent critical disorder
and the dependence of the calculated exponent on the particular 
average of $g$ used.   However, another possible factor is the properties 
of the leads for which the number of open channels represents an additional 
degree of freedom.  This requires further investigation.

\subsection{Scaling of the Lyapunov exponents\label{cube}}
For cubic systems attached to leads the contribution of the 1st
`Lyapunov exponent' \footnote{The Lyapunov exponents $\alpha_i$
are strictly defined only in the quasi--1D limit but in this work
the definition is extended to include arbitrary $L$ 
\eref{equ:lyan}.} to the value of the conductance in the critical
region can be calculated.  At $W=16.5$ the 1st LE makes up $85\%$
of the conductance and approximately $74\%$ of the derivative of
the conductance. Therefore the higher Lyapunov exponents do have
physical meaning even though the dominant contribution is from the
1st exponent. 
Scaling higher Lyapunov exponents is analogous to
scaling energy statistics at large energy scales (such as $P(i,s)$
for large $i$) \cite{Taylor98},
since both correspond to short length scale events.

\begin{figure}[htp]
    \begin{center}
        \mbox{\epsffile{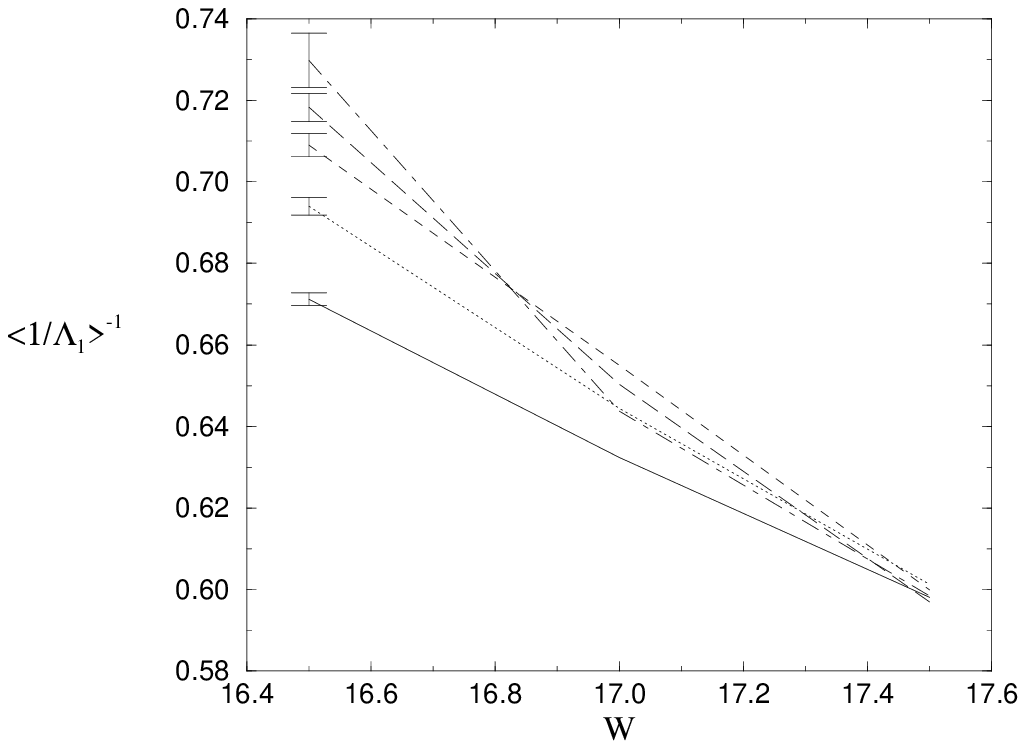}}\\
        \mbox{\epsffile{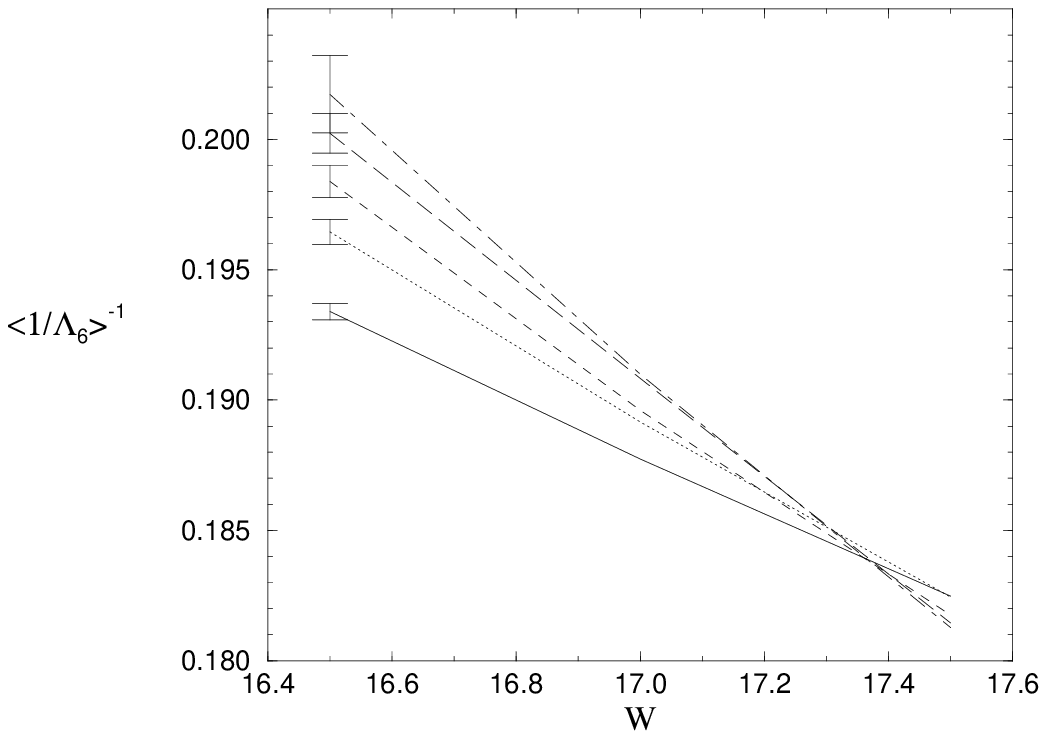}}
    \end{center}
    \caption{Dependence of $\left< \Lambda_1 \right>$ (upper figure) and 
$\left<\Lambda_6 \right>$ (lower figure) on disorder $W$ for
cubic systems of size $M$. Data points at $W=16.5,17.0,17.5$.  
$M=6$ (\full), $8$ (\dotted), $10$ (\broken), $12$ (\longbroken), 
\& $14$ (\chain).}
    \label{fig:lamcross}
\end{figure}

The 1st LE is plotted against disorder in
\fref{fig:lamcross}. Within the error bars of the calculation it
can be concluded that the critical point is above $W=16.5$ as was
the case with the conductance g. This shift of the critical point
is even clearer in the more accurate data for the 6th LE 
(\fref{fig:lamcross}). Qualitatively similar results occur for $\left<
\Lambda_i \right>$.

\begin{figure}[htp]
    \begin{center}
         \mbox{\epsffile{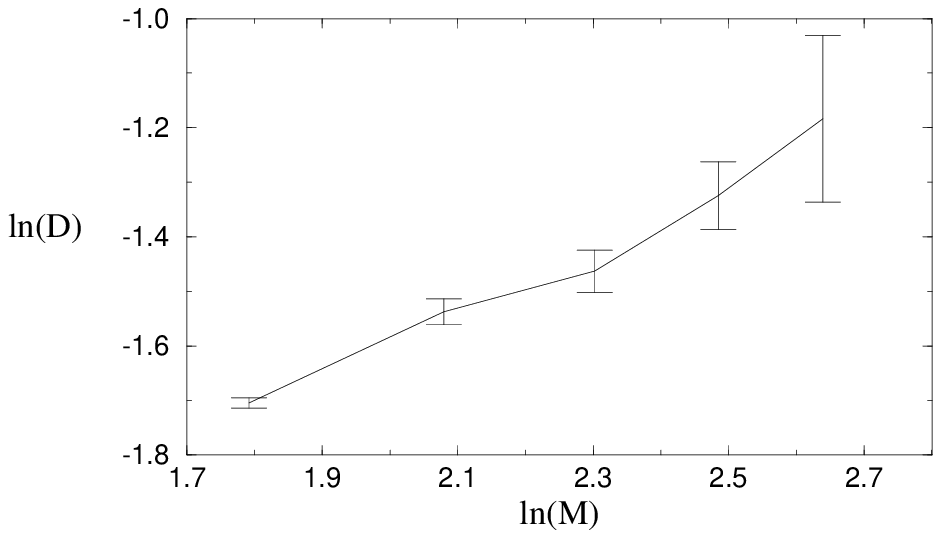}}\\
         \mbox{\epsffile{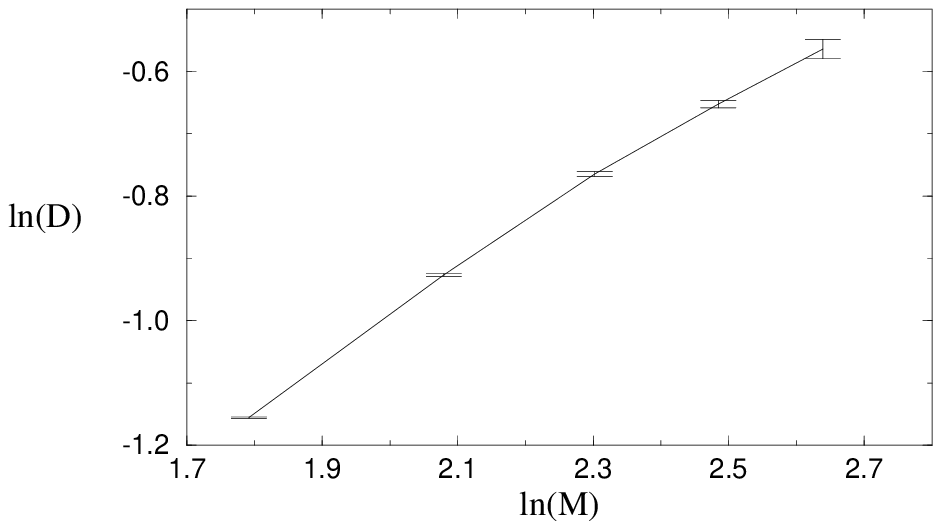}}
    \end{center}
    \caption{Scaling of the Lyapunov exponents. Slope is equal to $1/\nu$.
$D=\left< {\rmd\Lambda_i^{-1}/\rmd W} \right>$ at $i=1$ (upper figure), $i=6$ (lower figure).}
    \label{fig:scale1}\label{fig:scale6}
\end{figure}

In \fref{fig:scale1} equation \eref{equ:fit} is plotted with
$D=\left< {\rmd\Lambda_i^{-1}/\rmd W} \right>$ for $W=16.5$, i.e. the first
LE, $\left<\Lambda^{-1}\right>$, is being scaled.

\begin{figure}[htp]
    \center{\hspace{0cm}
         \epsffile{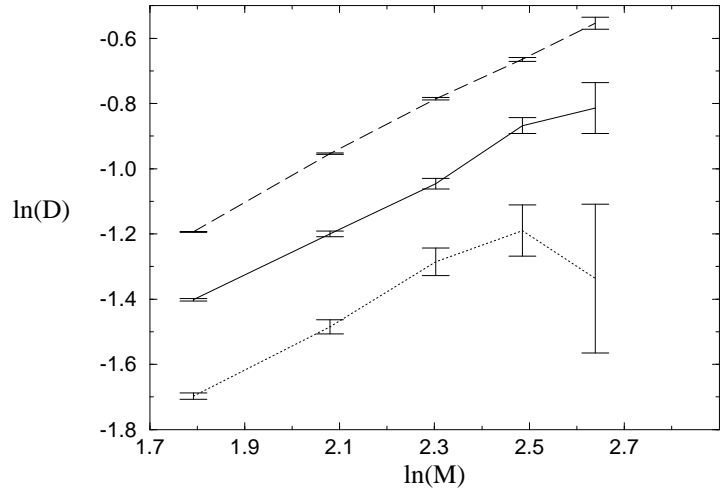}}
    \caption{Scaling of the 1st(\dotted), 2nd(\full) and 6th(\broken) Lyapunov exponents. Slope
is equal to $1/\nu$. $D=\left<{\rmd (1/ \Lambda_i)/\rmd W} \right>$, $W=17.5$}
    \label{fig:17p5scale}
\end{figure}

From one parameter scaling the gradient of this line is equal to
the inverse of the critical exponent.

\begin{figure}[htp]
    \begin{center}
        \ifcase\pdfoutput
           \mbox{\epsffile{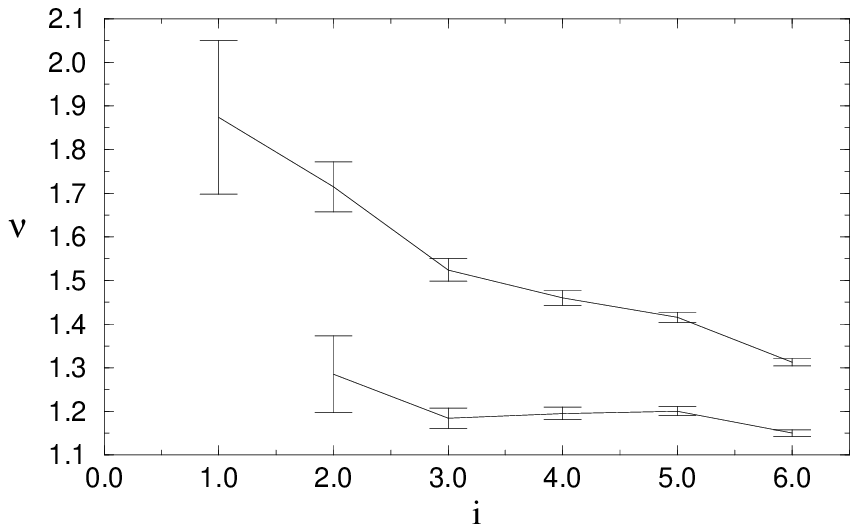}}\\
           \mbox{\epsffile{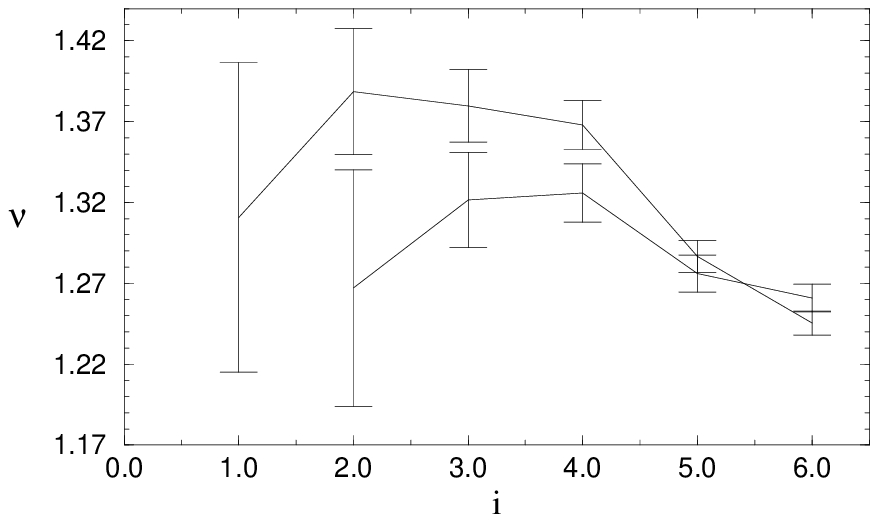}}
           \message{Outputting EPS Figures}
        \else
           \mbox{\pdfimage width 480 {16p5crit.pdf}}\\
           \mbox{\pdfimage width 480 {17p5crit.png}}
           \message{Outputting PDF Figures}
        \fi
        \end{center}
    \caption{Critical exponents calculated from $1/ \Lambda_i$ (higher curve)
and from $\Lambda_i$ (lower curve) for $i=1-6$. 
$W=16.5$ (upper figure), $W=17.5$ (lower figure)}
\label{fig:p5crit}
\end{figure}

The accuracy of the data increases when we consider higher LE's
and the result for the 6th LE is shown in \fref{fig:scale6}.
In theory this should be a straight line but the accuracy of the
data is high enough to observe a definite curvature indicating
deviations from one--parameter scaling theory. Although the curve
can be fitted by a straight line the quality--of--fit measure
$\chi^2$ is too large (and $Q \ll .1 $) which indicates the
curvature cannot be ignored. At $W=17.5$, which
is nearer the apparent critical point (\fref{fig:17p5scale}),
the results are qualitatively the same and the curvature in the $\ln(D)
\mathop{\rm vs} \ln(M)$ plot is again observed for the higher Lyapunov exponents.
This type of curvature indicates that scaling corrections exist
for the gradient.
A full analysis in terms of this equation was not possible because
the number of data points is too small to obtain sufficient
accuracy. In future studies with a full compliment of data from
$M=4$ to $14$ it is possible that the functional form of the
corrections could be found and subtracted to obtain a crossing at
$W=16.5$.
It is impossible to tell if the
curvature also exists for the 1st LE because it could be masked by
the larger error bars of the data.

The results of the straight line fit to calculate the critical
exponent are given in 
\fref{fig:p5crit}.
As mentioned, although the critical
exponents calculated from the higher Lyapunov exponents are more
accurate the $\chi^2$ becomes too large.
However the fits illustrate that the scaling behaviour is
approximately in agreement with previously calculated values of
the critical exponent.

\section{Conclusions}
We have introduced a new approach to calculating the critical behaviour
of the Anderson transition in disordered systems, which has the potential
to overcome some of the numerical difficulties which have plagued such
calculations in the past.  As the method is only requires a calculation
at a single value of disorder it eliminates the difficulties associated
with the necessity of fitting to a range of values of disorder and the
associated uncertainties associated with the determination of the range
of disorder over which the critical behaviour is valid.  The initial results
are promising, although the necessity of working with cubes rather than
long systems introduces deviations from the expected behaviour.  We expect
these deviations to be explicable in terms of the usual corrections to
scaling \cite{SO97,SO99,OSK99,SO01}.

\ack
The authors would like to thank Ben Simons, Derek Lee, Miles Blencowe, 
Keith Slevin  \& John Pendry for useful discussions.
APT would like to thank the EPSRC for a studentship. AMacK would like to
acknowledge the hospitality of the Cavendish Laboratory, Cambridge, where
this manuscript was completed.  

\section*{References}
\bibliography{disorder}

\end{document}